\documentclass[12pt]{article}
\input{epsf.sty}
\usepackage{hyperref}
\usepackage[pdftex]{graphicx}
\usepackage{amsmath, amssymb}
\def\phi{\varphi}

\def\T{\T}

\begin{document}
\title{Comment on the paper by \\ Y.~Komura and Y.~Okabe,
{\em J.Phys. A} {\bf 44}, 015002, 2011}

\author{
Aernout C.D. van Enter
\footnote{ University of Groningen,Johann Bernoulli Institute of Mathematics and
Computing Science, Postbus 407, 9700 AK Groningen, The
Netherlands,
\newline
\texttt{A.C.D.v.Enter@math.rug.nl},
\newline
\texttt{http://www.math.rug.nl/~aenter/ }}\, , Silvano Romano
\footnote{ University of Pavia, Dept of Physics, via A. Bassi 6, Pavia, Italy,   \newline
\texttt{Silvano.Romano@pv.infn.it},
\newline
\texttt{http://www.pv.infn.it/~romano/ }}\,
\, \\ and
Valentin A. Zagrebnov
\footnote{ Universit\'e de la M\'editerran\'ee and Centre de Physique
Th\'eorique - UMR 6207 Luminy Case 977, 13288 Marseille, Cedex09,  France
%\newline
\newline
\texttt \texttt{zagrebnov@cpt.univ-mrs.fr},
\newline
\texttt{http://www.cpt.univ-mrs.fr/~zagrebno/zagrebnov.htm}}}

\maketitle

\begin{abstract}
\noindent We point out that the claim of strong universality of \cite{KO} is incorrect, as it contradicts 
known rigorous results.

 \end{abstract}



\vfill\eject

The generalised lattice ferromagnetic $XY$-model was introduced by two of us in \cite{RZ}. Its simplest version involves
3--component unit spins on a $D$-dimensional lattice, parameterized by standard angles in the spherical coordinates and
interacting according to the ferromagnetic nearest-neighbour Hamiltonian
\begin{equation*}
H = -J\sum_{<i,j>} (sin \theta_i \, sin\theta_j)^q \ cos(\phi_i- \phi_j) \ , \ J > 0 \ .
\end{equation*}

When $D=2$, the model yields orientational disorder at all finite temperatures, and produces a transition to a
low--temperature Berezinski\v{\i}--Kosterlitz--Thouless (BKT) phase, possessing slow decay
of magnetic correlations and infinite susceptibility \cite{RZ}. In turn, the transition is expected, and was found by
simulation \cite{CPRM}, to exhibit the usual BKT scenario, at least for small values of $q$.
On the other hand, this scenario does not hold for \textit{all} values of $q$, i.e. the BKT transitions are \textit{not universal}.

Indeed, we \textit{proved} in \cite{ERZ} that the named transition turns first--order for sufficiently large $q$. Recall that a hint
of this behaviour was already suggested by numerical analysis \cite{CPRM}.

Notice that the above \textit{rigorous} results entail that both existence and the type of transition \textit{depend} on
the parameter $q$, i.e. they exclude universality.

The above-mentioned generalized $XY$--model was recently addressed by simulation for various values of $q$ in Ref. \cite{KO}.
Although the authors found only a certain evidence of the BKT criticality, they claim
that this behaviour holds for \textit{all values} of $q$.
This claim can not be correct since it contradicts to our rigorous results  \cite{ERZ}, which seem to have
been unknown to the authors of the Ref. \cite{KO}.


\begin{thebibliography}{99}

\bibitem{KO} Y.~Komura and Y. Okabe,  {\em J.Phys. A} {\bf 44}, 015002, 2011.

\bibitem{RZ} S.~Romano and V.~A.~Zagrebnov, {\em Phys. Lett. A} {\bf 301}, 201, 2002.

\bibitem{CPRM} L.~A.~S. Mol, A.~R.~ Pereira, H.~Chamati  and S.~Romano, {\em Eur.Phys. J. B} {\bf 50}, 541, 2006.

\bibitem{ERZ} A.~C.~D. van Enter, S.~Romano and V.~A.~Zagrebnov, {\em J. Phys. . A} {\bf 39}, L439, 2006.

\end{thebibliography}
\end{document}